\documentclass{eas}
\usepackage{graphicx}
\begin{document}

\title{Metallicity Effects in PDRs} 
\author{Markus R\"ollig}\address{Argelander Institut f\"ur Astronomie, Universit\"at Bonn, Germany\\
now: 1.Physikalisches Institut, Univ. zu K\"oln, Germany}
%
%
\begin{abstract}
Almost all properties of a photodissociation region (PDR) depend on its metallicity. The heating and cooling efficiencies that determine the temperature of the gas and dust, the dust composition,  as well as the elemental abundances that influence the chemical structure of the PDR are just three examples that demonstrate the importance of metallicity effects in PDRs. PDRs are often associated with sites of star formation. If we want to understand the star formation history of our own Galaxy and of distant low-metallicity objects we need to understanding how metallicity acts on PDR physics and chemistry.
\end{abstract}
\maketitle
\section{Introduction}
\subsection{PDRs and Star Formation}
It is common to define PDRs as regions where stellar far ultraviolett (FUV: 6~eV$ \le h\nu \le$ 13.6~eV) radiation dominates the physical and chemical properties of the local interstellar medium (ISM). FUV radiation is predominantely produced in massive stars, which do not live long enough to exit their parental cloud of gas and dust within their lifetime. PDRs are the products of the strong mutual feedback between young massive stars and their parental clouds and hence closely related to the process of star formation (SF). Metallicity ($Z$) is thought to affect the fragmentation properties of gravitationally unstable gas, especially in the early universe (Bromm {\em et al.\/} \cite{bromm01}; Schneider {\em et al.\/} \cite{schneider02}). This should also affect the initial mass function (IMF). However, Jappsen {\em et al.\/} \cite{jappsen07} have shown, that the cooling of hot, very low-$Z$ gas does not seem not to depend on metallicity, since it is dominated by H$_2$ cooling. Hence, the cooling depends on the amount of H$_2$ available for cooling, which might still be a function of metallicity due to the fact that molecular hydrogen is formed on metallic surfaces, i.e. dust.   
\subsection{Metallicity}
 \begin{figure}
 \centering
 \includegraphics[width=0.8\columnwidth]{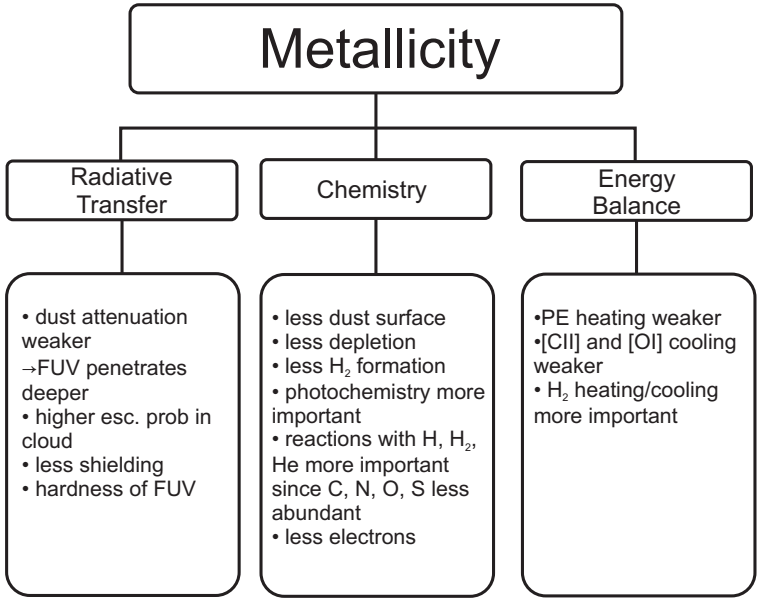}
\caption{Overview of metallicity influence on different PDR aspects.}\label{influence}
\end{figure}
\begin{figure}
 \centering
 \includegraphics[width=0.6\columnwidth]{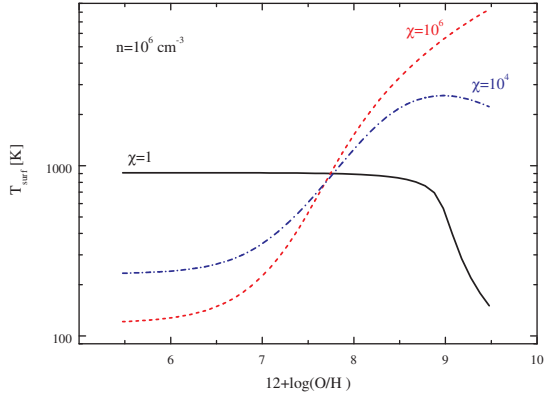}
\caption{Surface gas temperature of a high density PDR vs. metallicity for different FUV intensities. At low $Z$ an inversion occurs, i.e. the surface temperature drops for an intenser FUV illumination.}\label{invertedT}
\end{figure}

Figure~\ref{influence} lists the most prominent metallicity effects. An altered metallicity affects the physics (primarily energy balance and radiative transfer) and the chemistry of PDRs. The most important, direct effects are altered elemental abundances leading to a different chemical structure and different total column densities of important species like C$^+$, C, and CO as well as an altered dust-to-gas ratio $D/G$. A lower $D/G$ leads to a weaker attenuation and FUV radiation penetrates deeper into the cloud, destroying fragile molecule up to larger depths. Additionally,  gas heating by FUV photons can act deeper in the cloud leading to a different temperature structure compared to solar metallicity PDRs. On the other hand, the probability for a photon to escape the cloud is increased, improving the cooling efficiency at any particular depth while the smaller elemental abundance reduces the cooling rate. Observations of low metallicity environments also confirm that the dust composition changes at lower $Z$. The PAH (polycyclic aromatic hydrocarbons) fraction drops significantly (Draine {\em et al.\/} \cite{draine07}, Madden {\em et al.\/} \cite{madden06}). PAHs dominantly contribute to the total photoelectric heating (PE) efficiency, thus, at lower values of $Z$ the PE efficiency is significantly diminished. Madden {\em et al.\/} \cite{madden06} also showed, that the spectral shape of the FUV radiation is different in low $Z$ dwarf galaxies. Reduced metallicities lead,  for a given stellar type, to hotter main sequence temperatures and
to harder, i.e. more energetic, spectral SEDs. This may affect the dust composition via evaporation and photodissociation processes and leads to a different PE efficiency. Since all these effects are non-linear and mutually coupled it is very difficult to give an analytical solution and one has to rely on numerical computations.         
Sometimes the results are completely counter-intuitive like displayed in Figure~\ref{invertedT}. The heating and cooling efficiencies at the cloud surface depend on parameters like density $n$, metallicity and FUV intensity $\chi$.
With increasing FUV intensities and densities above $\sim10^4$~cm$^{-3}$, the dominant heating contribution shifts from H$_2$-deexcitation to  photoelectric (PE) heating (c.f. R\"ollig  {\em et al.\/} \cite{roellig06}).
Since $\Gamma_\mathrm{PE}\sim Z^2$ is much less effective for $Z<1$ compared to the cooling terms, which scale linearly with $Z$, the gas temperature drops if $\chi$ increases, i.e. if we move closer to the FUV source! Vice versa this means, a hypothetical cloud gets hotter the further away it is placed from a strong FUV source. Even though this might be an academic example due to its particular assumptions, it demonstrates the difficulty in predicting metallicity effects.  
\subsection{The KOSMA-$\tau$ PDR Code}
The model results presented here are computed with the KOSMA-$\tau$ PDR code. (Stoerzer  {\em et al.\/} \cite{stoerzer96}, R\"ollig  {\em et al.\/} \cite{roellig06}). This code features spherical geometry and an isotropical FUV irradiation. It calculates the chemical and physical cloud structure as function of radius as well as the total cloud emission. The metallicity is a free model parameter and changes the elemental abundances and the gas-to-dust ratio. The chemical network in KOSMA-$\tau$ is fully modular and individual species can be easily added to or removed from the network. Recently, the code has been expanded to calculate the dust temperatures and continuum emission for any given dust distribution (R\"ollig \& Szczerba \cite{roellig08}).  The model setup can be easily expanded to clump ensembles as presented by Cubick {\em et al.\/} (this issue) to calculate the PDR emission of a large, complex superposition of differently sized clouds. R\"ollig  {\em et al.\/} \cite{roellig07} gives an overview over PDR modelling as well as a comparison between a number of different model codes. The KOSMA-$\tau$ results are 
also available online ({\small \tt http://www.astro.uni-koeln.de/workgroups/theo\_astronomy/pdr/}).
\section{Shielding and PDR Structure}
 \begin{figure}
 \centering
 \begin{minipage}[hbt]{0.49\columnwidth}
 \centering
  \includegraphics[width=\columnwidth]{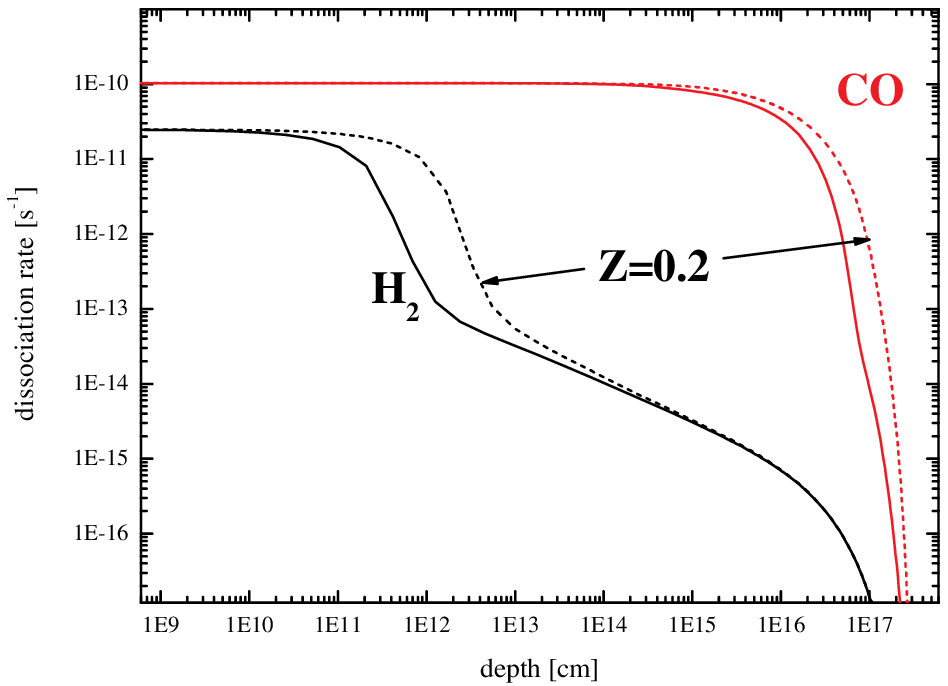}
 \end{minipage}
\hfill
 \begin{minipage}[hbt]{0.49\columnwidth}
 \centering
 \includegraphics[width=\columnwidth]{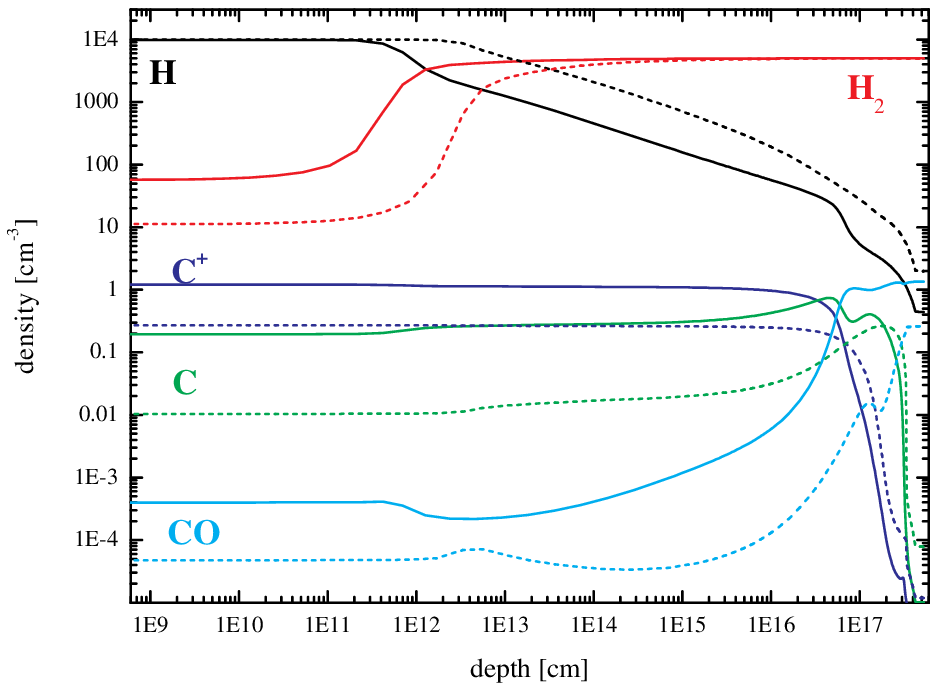}
\end{minipage}
\caption{Depth dependent photodissociation rates and chemical abundances for different metallicities.  Solid lines are for solar metallicities, dashed lines correspond to $Z=0.2$. {\bf Left:} Photodissociation rates of H$_2$ (black) and CO (red) vs. depth [cm]. {\bf Right:} Chemical abundances vs. depth. The lower metallicity leads to a lower abundance at the surface of the cloud. Additionally, the weaker shielding shifts the typical H-H$_2$ and C$^+$-C-CO transitions deeper into the cloud.}\label{shielding}
\end{figure}
A reduced metallicity leads to a decreased shielding of FUV radiation. In the left panel in Figure~\ref{shielding} we compare the H$_2$ and CO photodissociation rates for $Z=0.2$ and $Z=1$. The right panel shows the chemical abundances vs. depth. The lower metallicity leads to a lower abundance of all metallic species at the surface of the cloud. Additionally, the weaker shielding shifts the typical H-H$_2$ and C$^+$-C-CO transitions deeper into the cloud leading to an almost negligible amount of CO in the $Z=0.2$ case while the C$^+$ column depth is much less affected.   
\section{PDR Emission}
\begin{figure}
 \includegraphics[width=0.48\columnwidth]{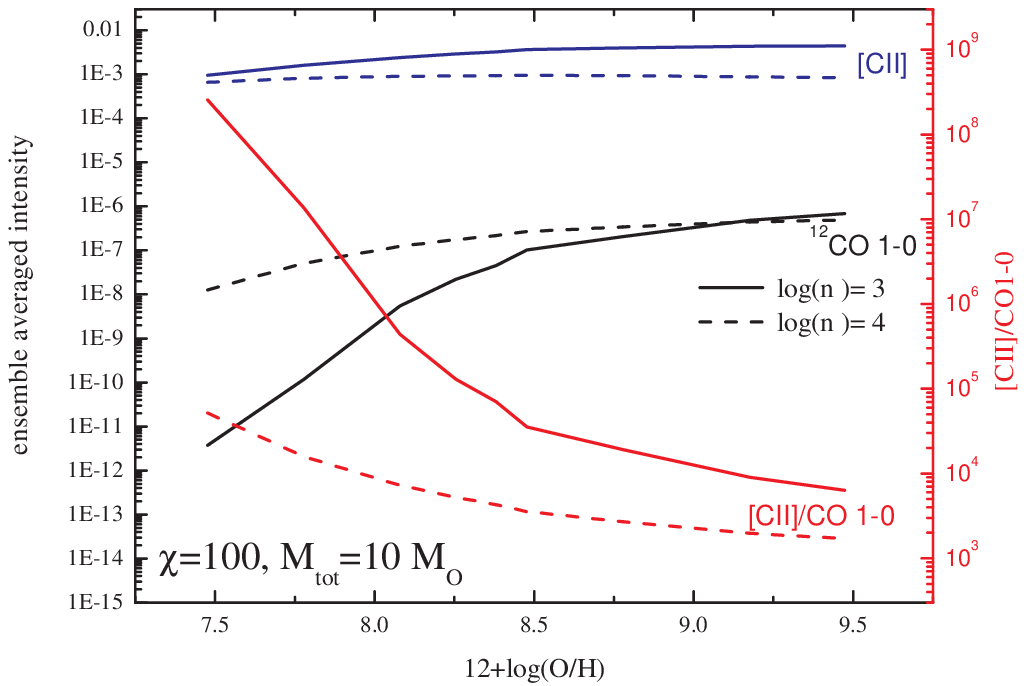}
\qquad
 \includegraphics[width=0.45\columnwidth]{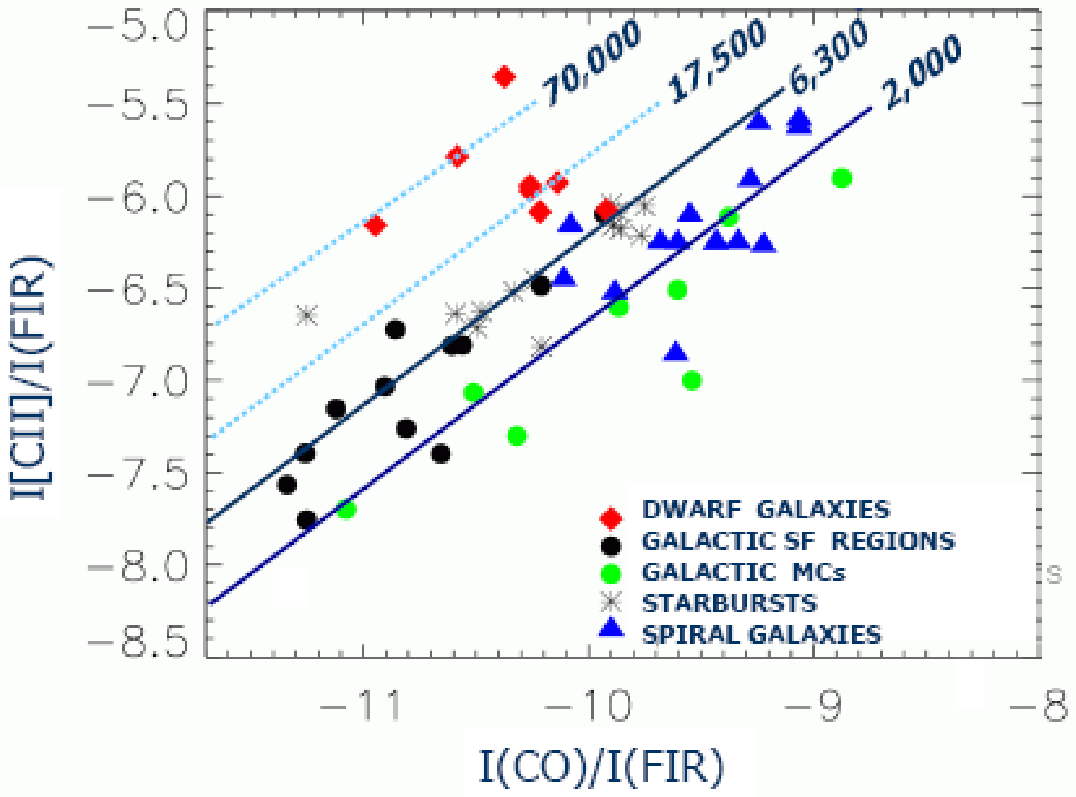}
\caption{{\bf Left:} [CII] and CO(1-0) surface brightnesses of an ensemble of model PDRs vs. metallicity. The red lines show the ratio [CII]/CO(1-0). {\bf Right:}[CII] survey results: comparing normal metallicity regions with
low-metallicity galaxies (from Madden \cite{madden00}).}\label{CIIvsCO}
\end{figure}
The reduced column densities of the more fragile species at lower metallicities are visible as lower surface brightnesses of the PDR. CO, being one of the most important chemical species, suffers strongly from the weak shielding due to low $Z$. Figure~\ref{CIIvsCO} compares the ensemble averaged clump emission (see below) of [CII], and CO(1-0) vs. metallicitiy (on the 12+log(O/H) scale). For lower metallicity values, the CO surface brightness decreases significantly. This is a combination of reduced column densities plus a much smaller projected area of the CO core of the cloud due a transition from C to CO at much higher depths. This projection effect is particularly strong in the low density case $n=10^3$~cm$^{-3}$, visible as kink in the CO curve at 12+log(O/H)=8.4.  The right panel in Figure \ref{CIIvsCO}, taken from Madden \cite{madden00}, shows the [CII] intensity vs. CO(1-0) comparing normal metallicity regions with low-metallicity galaxies. Lines of constant $I_\mathrm{[CII]}/I_\mathrm{CO}$ run diagonally across the plot. The Figure clearly shows the metallicity dependence of the [CII]/CO ratio. A similar behavior has been found by Nakagawa {\em et al.\/} \cite{nakagawa99} in low metallicity clouds in the Large Magellanic Clouds (LMC).     
\section{Clumpy Composition of Low-Z PDRs}
The ratio $I_\mathrm{[CII]}/I_\mathrm{CO}$ is a good measure of PDR emission relative to the emission of the cold molecular material and is a good indicator  of star formation. Active star formation in the Milky Way and in nearby star burst galaxies shows a ratio $I_\mathrm{[CII]}/I_\mathrm{CO}=6300$ while dwarf galaxies show ratios between 6000 and 70000. This makes dwarf galaxies one of the prime targets to study low metallicity PDRs. Due to their small angular size, single dish far-infrared (FIR) observations of dwarf galaxies are unable to resolve individual molecular clouds. Hence we apply a clump ensemble approach (Cubick {\em et al.\/} \cite{cubick08}) to model the beam averaged PDR emission of these galaxies.  
 \begin{figure}
 \centering
 \includegraphics[width=1\columnwidth]{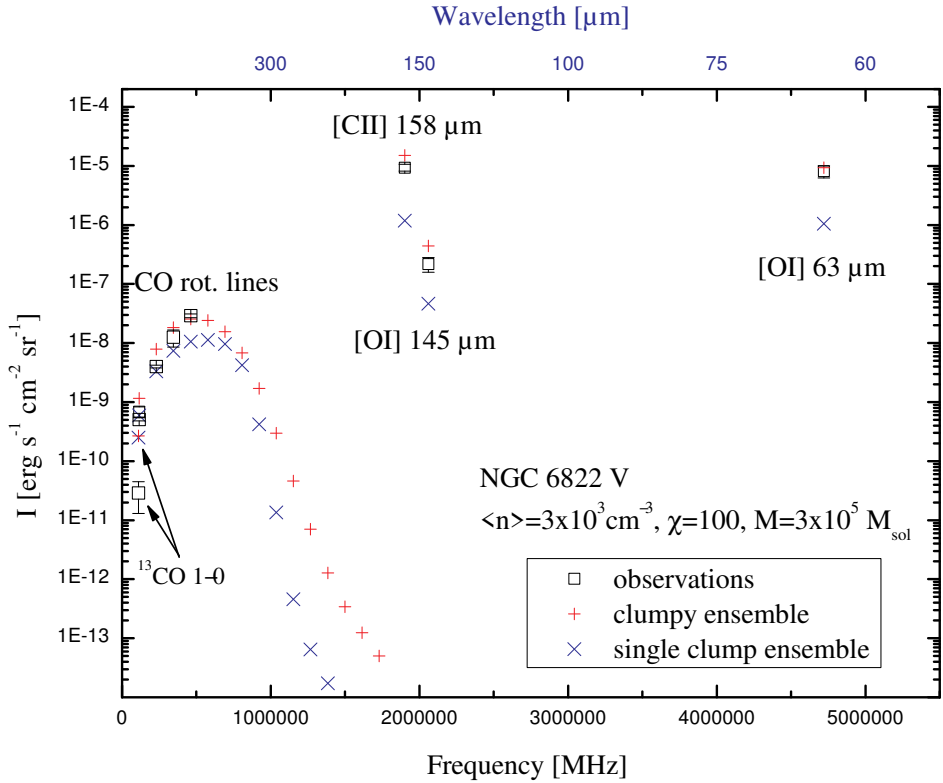}
\caption{PDR Model results for NGC~6822V. The red crosses indicate the model values for a clump ensemble while the blue crosses show the results for a single component PDR model. CO rotational lines are in the lower left, fine-structure cooling lines in the upper right. The open squares and circles denote observed data points (Israel {\em et al.\/} \cite{israel03}, Madden, priv. comm.).}\label{ngc6822}
\end{figure}
Figure~\ref{ngc6822} compares the model results for the PDR emission of NGC~6822V, an HII region complex in the dwarf irregular galaxy of the local group at a distance of 470~kpc and a $Z\approx 0.82\,Z_\odot$ (12+log(O/H)=8.37), with FIR observations. The blue crosses in Figure~\ref{ngc6822} show the results for a single component model, i.e. the galaxy is assumed to consists of numerous identical model clouds. The red crosses denote a clump-mass distribution  $dN/dM\sim M^{-1.8}$ and  mass-size relation $M\sim R^{2.3}$. Taking into account the observed mid-$J$ CO and fine-structure transitions it is not possible to find a good model fit for a single component model. For a clump ensemble we derive a mean gas density $<n>=3\times 10^3$~cm$^{-3}$, $\chi=100 \chi_\mathrm{Draine}$, $M_\mathrm{gas}=3\times 10^5\,M_\odot$.


\end{document}